\documentclass[aps,reprint,prl,longbibliography,nofootinbib,amsmath,amssymb,amsfonts,notitlepage]{revtex4-1}
\usepackage[utf8]{inputenc}
\usepackage{graphicx}
\usepackage{amsfonts,amsthm}
\usepackage{bm}
\usepackage[usenames,dvipsnames]{xcolor}

\newcommand{\be}{\begin{equation}}
\newcommand{\ee}{\end{equation}}
\newcommand{\bea}{\begin{eqnarray}}
\newcommand{\eea}{\end{eqnarray}}
\renewcommand{\Re}{\mathrm{Re }}

\date{\today}

\newcommand{\dst}{\displaystyle}
\newcommand{\lr}[1]{ \langle #1 \rangle}
\newcommand{\fr}[2]{\frac{{\dst #1}}{{\dst #2}}}
\renewcommand{\Re}{\mathrm{Re }}

\newcommand{\mmmatrix}[9]{ \left(\! \begin{array}{ccc}#1 & #2 & #3\\[1mm] #4 & #5 & #6\\[1mm] #7 & #8 & #9\\ \end{array}\!\right) }
\newcommand{\doublet}[2]{ \left( \begin{array}{c}#1 \\ #2 \end{array}\right) }
\newcommand{\triplet}[3]{ \left( \begin{array}{c}#1 \\[2mm] #2 \\[2mm] #3\end{array}\right) }

\newcommand{\bA}{{\bf A}}

\newcommand{\bbe}{{\bf e}}
\newcommand{\bk}{{\bf k}}
\newcommand{\bK}{{\bf K}}

\newcommand{\br}{{\bf r}}

\def\lsim{\mathrel{\rlap{\lower4pt\hbox{\hskip1pt$\sim$}}
    \raise1pt\hbox{$<$}}}         
\def\gsim{\mathrel{\rlap{\lower4pt\hbox{\hskip1pt$\sim$}}
    \raise1pt\hbox{$>$}}}         

    \usepackage{bbold}

\newcommand{\ga}{\gamma}
\newcommand{\de}{\delta}

\newcommand{\la}{\lambda}

\newcommand{\si}{\sigma}

\begin{document}
	\title{
		{\normalsize \hfill CFTP/19-030} \\*[7mm]
		Doing spin physics with unpolarized particles}
		\author{Igor P. Ivanov}
	\email{igor.ivanov@tecnico.ulisboa.pt}
	\affiliation{CFTP, Instituto Superior Tecnico, Universidade de Lisboa, Lisbon 1049-001, Portugal}
		\author{Nikolai Korchagin}
    \email{korchagin@impcas.ac.cn}
    \affiliation{Institute of Modern Physics, Chinese Academy of Sciences, Lanzhou 730000, China}
		\author{Alexandr Pimikov}
    \email{pimikov@mail.ru}
    \affiliation{Institute of Modern Physics, Chinese Academy of Sciences, Lanzhou 730000, China}
    \affiliation{Research Institute of Physics, Southern Federal University, Rostov-na-Donu 344090, Russia}
		\author{Pengming Zhang}
    \email{zhangpm5@mail.sysu.edu.cn}
    \affiliation{School of Physics and Astronomy, Sun Yat-sen University, Zhuhai 519082, China}

\begin{abstract}
Twisted, or vortex, particles refer to freely propagating non-plane-wave states 
with helicoidal wave fronts. In this state, the particle possesses a non-zero orbital angular momentum
with respect to its average propagation direction.
Twisted photons and electrons have been experimentally demonstrated, 
and creation of other particles in twisted states can be anticipated. 
If brought in collisions, twisted states offer a new degree of freedom to particle physics,
and it is timely to analyze what new insights may follow. 
Here, we theoretically investigate resonance production 
in twisted photon collisions and twisted $e^+e^-$ annihilation
and show that these processes emerge as a completely novel probe of spin and parity-sensitive observables
in fully inclusive cross sections with unpolarized initial particles.
This is possible because the initial state with a non-zero angular momentum
explicitly breaks the left-right symmetry even when averaging over helicities.
In particular, we show how one can produce almost $100\%$ polarized vector mesons
in unpolarized twisted $e^+e^-$ annihilation and how to control its polarization state. 
\end{abstract}
	
	\maketitle

{\em Introduction.}
Spin-parity properties of hadrons are a fascinating chapter in modern particle phenomenology.
The rich hadron spectrum exhibits a variety
of spin-parity quantum numbers possible in the $q\bar q$ and $qqq$ quark combinations 
and in multiquark states \cite{Olsen:2014qna}.
Deep inelastic scattering (DIS) with polarized lepton or proton
allows one to investigate how spin of the ultrarelativistic proton emerges
from spins and orbital angular momenta of its constituents \cite{Aidala:2012mv,Leader:2013jra}.
Reconstructing the proton spin structure in 3D brings in new spin-sensitive variables, 
which can be encoded via transverse momentum distributions and explored experimentally
in semi-inclusive DIS with transversely polarized protons \cite{Anselmino:2007fs}.

Spin-parity structure of hadrons is explored in experiment via two approaches:
either one collides polarized initial particles
and investigates the response of the cross section to flipping the polarization sign or direction,
or one studies exclusive or semi-inclusive reactions
and reconstructs the spin properties of the target hadron or intermediate resonances
from the final state angular distributions.
There seems to be no other way to access spin-dependent observables.

Here, we propose a novel tool for spin physics in particle collisions, 
complementary to all existing approaches. 
We demonstrate that parity- and spin-sensitive observables can be accessed
in fully inclusive processes with unpolarized initial particles --- provided they 
are prepared in the so-called {\em twisted} states.
This initial state explicitly breaks the left-right symmetry even when averaging over the initial particle helicities.
We calculate resonance production in unpolarized twisted $\gamma\gamma$ collision or in twisted $e^+e^-$ annihilation
and demonstrate that such processes offer unprecedented control over 
polarization of the produced resonance.

{\em Twisted photons and electrons.}
A twisted particle, be it a photon, an electron, or a hadron, 
is a wave packet with helicoidal wave fronts.
It propagates, as a whole, in a certain direction and carries 
a non-zero orbital angular momentum (OAM) projection with respect to that direction, which can be adjusted experimentally.
Twisted photons are known since long ago \cite{Allen:1992zz,Molina-Terriza:2007,Paggett:2017,Knyazev:2019};
following suggestions of \cite{Bliokh:2007ec}, twisted electrons \cite{Bliokh:2017uvr,Lloyd:2017} were also recently created 
\cite{Uchida:2010,Verbeeck:2010,McMorran:2011}. 
Relying on ideas of how to bring them into the GeV energy range \cite{Jentschura:2010ap,Jentschura:2011ih} and
steer them in accelerators \cite{Silenko:2019ziz},
and on future experimental progress, one can imagine collider-type experiments with
GeV-range twisted photons or twisted $e^+e^-$ annihilation. 

Description of twisted particles adapted to calculation of their high energy collisions 
was presented in \cite{Jentschura:2010ap,Jentschura:2011ih} 
and further developed in \cite{Ivanov:2011kk,Ivanov:2011bv,Karlovets:2012eu},
see also recent reviews \cite{Bliokh:2017uvr,Knyazev:2019} and the Supplementary information.
For each sort of fields (photons, electrons, etc), 
one begins with Bessel twisted states $|E,\varkappa,m\rangle$, 
which represent solutions of the free wave equation 
with definite energy $E$, longitudinal momentum $k_z$, 
modulus of the transverse momentum $|\bk_\perp|=\varkappa$,
the $z$-projection of the total angular momentum (AM) $m$, 
and definite helicity. 
(We use natural units $\hbar = c = 1$ and denote three-dimensional vectors by bold symbols,
labeling the transverse momenta with $\perp$.)
A Bessel twisted photon is defined, in the Coulomb gauge, as 
\begin{equation}
\bA_{\varkappa m k_z \lambda}(\br) = e^{i k_z z} \int a_{\varkappa m}(\bk_\perp)\, \bbe_{\bk \lambda}\, e^{i\bk_\perp \br_\perp} {d^2\bk_\perp \over (2\pi)^2}\,,
\label{tw1-ph}
\end{equation}
where the Fourier amplitude $a_{\varkappa m}(\bk_\perp)$ is given by
\begin{equation}
a_{\varkappa m}(\bk_\perp) = i^{-m} e^{im\varphi_k} \sqrt{2\pi \over \varkappa}\delta(|\bk_\perp| - \varkappa)\,.\label{a-ph}
\end{equation}
It is a superposition of plane wave (PW) photons with equal $E$, $k_z$,
$|\bk_\perp| = k\sin\theta$, and helicity $\lambda = \pm 1$, but arriving from different azimuthal angles $\varphi_k$.
Each PW component of a twisted photon contains its polarization vector $\bbe_{\bk \lambda}$,
orthogonal to its momentum: $\bbe_{\bk\lambda} \bk = 0$.
Notice that a twisted photon with given total AM $m$ and helicity $\lambda$ is not an eigenstate 
neither of the OAM $z$-component operator 
$\hat{L}_z = -i \partial/\partial \varphi_k$ nor of the spin $z$-component operator $\hat{s}_z$.
This is a manifestation of the spin-orbital interaction of light,
which gives rise to a variety of remarkable optical phenomena \cite{Bliokh:2015yhi}.
Nevertheless, in the paraxial approximation $\varkappa/|k_z| = |\tan\theta| \ll 1$, 
the spin-orbital coupling is suppressed and one can deal with approximately conserved 
$s_z \approx \lambda$ and $L_z \approx \ell \equiv m - \lambda$.

Similarly, a Bessel twisted electron
\cite{Bliokh:2011fi,Karlovets:2012eu,Bliokh:2017uvr}
as a monochromatic solution of the Dirac equation with definite $E$, $k_z$, $\varkappa$,
half-integer total AM $m$, and helicity $\zeta = \pm 1/2$:
\be
\label{bessel}
\Psi_{\varkappa m k_z \zeta}(\br) = e^{i k_z z} 
\int a_{\varkappa m}(\bk_\perp)\, {u_{\zeta}(k) \over \sqrt{2E}} \, e^{i\bk_\perp \br_\perp} {d^2\bk_\perp \over (2\pi)^2}\,,
\ee
with the standard expressions for the plane wave bispinor $u_{\zeta}(k)$ and 
with the same Fourier amplitude $a_{\varkappa m}(\bk_\perp)$ as in Eq.~\eqref{a-ph}.
Similar expression holds for the negative frequency solutions of the Dirac equation $v_{\zeta}(k)$ 
used to describe positrons. Again, spin and OAM $z$-projections 
are not separately conserved in twisted electron due to the intrinsic spin-orbital interaction 
\cite{Bliokh:2011fi,Karlovets:2018iww},
but in the paraxial approximation 
both $s_z \approx \zeta$ and $\ell = m - \zeta$ are approximately conserved in the focal spot.

A pure Bessel state $|E,\varkappa,m\rangle$ 
is not normalizable in the transverse plane.
We construct a realistic normalizable monochromatic twisted beam as a superposition of Bessel states 
with equal $E$, $m$, and helicity but with a distribution over $\varkappa$,
\begin{equation}
|E,\bar\varkappa,\sigma,m\rangle = \int d\varkappa \, f(\varkappa) |E,\varkappa,m\rangle\,,\label{WP}
\end{equation}
with a weight function $f(\varkappa)$ peaked at $\bar\varkappa$ and having a width $\sigma$. 

{\em Unpolarized twisted photons or electrons.}
For PW initial particles, the polarization and coordinate degrees of freedom factorize,
and the definition of the unpolarized cross section is straightforward. 
For twisted photons and electrons, where they are coupled, this notion requires clarification.
When averaging cross section over initial helicities, should one keep their $m$ unchanged or not?

In fact, there is no unique definition of unpolarized twisted photon or electron beam, 
because it will eventually depend on the details of experimental realization.
If an experimental device is capable of selecting electrons
in a single $m$ state irrespective of $\zeta$,
then one calculates the cross sections 
$\sigma_+$ with the twisted state $|m, \zeta = +1/2\rangle$ 
and $\sigma_-$ with the twisted state $|m, \zeta = -1/2\rangle$
and obtains the unpolarized cross section as $\sigma_0 = (\sigma_+ + \sigma_-)/2$.
These $\sigma_+$ and $\sigma_-$ are not expected to be always equal.
Indeed, although $|m, \zeta = +1/2\rangle$ is not an OAM eigenstate,
it is dominated by the OAM component $\ell = m-1/2$.
Similarly, the state $|m, \zeta = -1/2\rangle$ is dominated by $\ell' = m+1/2 = \ell + 1$.
These states have different current and spin densities \cite{Bliokh:2010pc,Bliokh:2011fi,Karlovets:2018iww}.
Therefore, they may lead to non-equal collision cross sections even 
if the fundamental interactions are $P$-invariant.

Another possible definition of an unpolarized twisted beam 
is to keep $\ell = m - \zeta$ fixed. 
However since a twisted beam is not an OAM eigenstate,
its downstream evolution leads to spin-to-orbital conversion, which was experimentally 
verified for twisted light \cite{beads2007}.
In real experiments, the sensitivity of the cross section to the helicity flip 
will depend on experimental set-up and cannot be predicted unambiguously. 
However, the effect will be there for any setting.
To elucidate its particle physics consequences, 
we use below the fixed-$m$ definition and demonstrate
that twisted particle annihilation is a remarkably rich tool for spin physics. 


{\em Resonance production in twisted particle collisions.}
We apply the formalism developed in \cite{Jentschura:2010ap,Ivanov:2011bv} to a generic $2\to 1$ process
of resonance production in annihilation of two counter-propagating 
twisted initial particles defined with respect to the same axis $z$
with energies $E_1$, $E_2$, moduli of the transverse momenta $\varkappa_1$, $\varkappa_2$, 
and AM $m_1$ and $m_2$, respectively. For simplicity, the two initial particles are assumed to be massless.
The final particle with mass $M$ is described in terms of plane waves with momentum $\bK$ and energy $E_K$.

The $S$-matrix amplitude of twisted particle annihilation can be written as a superposition
of PW $S$-amplitudes $S_{PW}$ with different transverse momenta of the initial particles:
\begin{eqnarray}
S &=& \int {d^2 \bk_{1\perp} \over (2\pi)^2} {d^2 \bk_{2\perp} \over (2\pi)^2} 
a_{\varkappa_1 m_1}(\bk_{1\perp}) a_{\varkappa_2, -m_2}(\bk_{2\perp}) S_{PW} \nonumber\\
&=& 
i (2\pi)^4 \fr{\delta(\Sigma E) \delta(\Sigma k_z)}{\sqrt{8 E_1 E_2 E_K}} 
{(-i)^{m_1-m_2} \over (2\pi)^3\sqrt{\varkappa_1\varkappa_2}} \cdot {\cal J}\,,\label{S-tw5}
\end{eqnarray}
where $\delta(\Sigma E) \equiv \delta(E_1+E_2-E_K)$, $\delta(\Sigma k_z) \equiv \delta(k_{1z}+k_{2z}-K_z)$.
The twisted amplitude ${\cal J}$ is defined as
\begin{equation}
{\cal J} = \varkappa_1\varkappa_2 \!\! \int\!\! d\varphi_1 d\varphi_2\, e^{im_1\varphi_1 - im_2\varphi_2}\,
\delta^{(2)}(\bk_{1\perp}+\bk_{2\perp} - \bK_\perp)\cdot {\cal M}\,,\label{J}
\end{equation}
where ${\cal M}$ is the usual PW invariant amplitude. 
\begin{figure}[ht]
\centering
\includegraphics[width=0.5\textwidth]{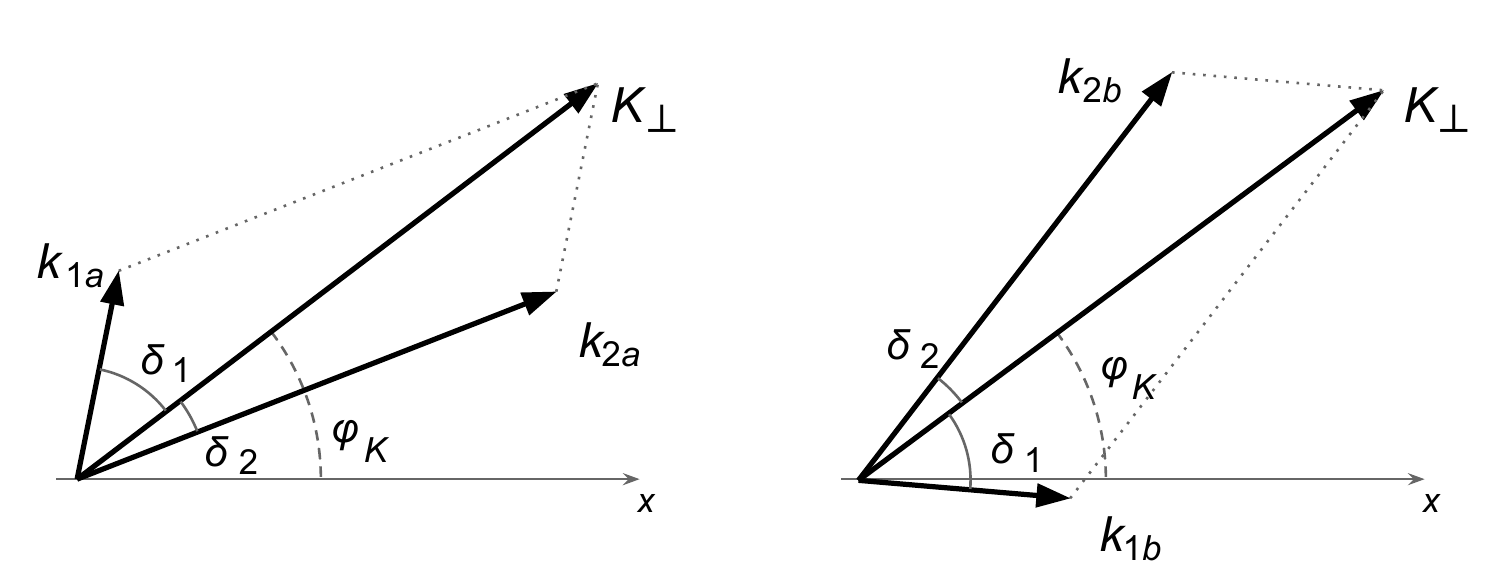}
{\caption{\label{fig-2configurations} The two kinematic configurations in the transverse plane that satisfy
momentum conservation laws in scattering of two Bessel states.}}
\end{figure}
The twisted amplitude ${\cal J}$ is nonzero only if $\varkappa_i = |\bk_{i\perp}|$ and $K \equiv |\bK_\perp|$ 
satisfy the triangle inequalities
\be
|\varkappa_1 - \varkappa_2| \le K \le \varkappa_1 + \varkappa_2\label{ring}
\ee
and form a triangle with the area $\Delta$. 
Out of the many PW components ``stored'' in the initial twisted particles, 
the integral \eqref{J} receives contributions from exactly two plane wave combinations shown in Fig.~\ref{fig-2configurations}:
(a) $\varphi_1 = \varphi_{K} + \delta_1$, $\varphi_2 = \varphi_{K} - \delta_2$,
(b) $\varphi_1 = \varphi_{K} - \delta_1$, $\varphi_2 = \varphi_{K} + \delta_2$,
where $\delta_1$ and $\delta_2$ are the inner angles of the triangle.
As a result, ${\cal J}$ can be calculated exactly \cite{Ivanov:2011kk}: 
\begin{equation}
	{\cal J} \propto {\varkappa_1 \varkappa_2 \over 2\Delta}
\left[{\cal M}_{a}\, e^{i (m_1 \delta_1 + m_2 \delta_2)} + {\cal M}_{b}\, e^{-i (m_1 \delta_1 + m_2 \delta_2)}\right]\,.\label{J2}
\end{equation}
One observes the hallmark feature of twisted particle collisions:
interference between two PW amplitudes ${\cal M}_{a}$ and ${\cal M}_{b}$ 
calculated for the two distinct initial PW pairs shown in Fig.~\ref{fig-2configurations} 
but the same final momentum $\bK_\perp$.

The cross section can be written as 
\be
d\sigma \propto |{\cal J}|^2 \delta(E_1+E_2-E_K) \, d^2 \bK_\perp\,.\label{dsigma-tw}
\ee
The prefactor here is inessential since the new effects come not from the overall magnitude of the cross section
but from its kinematic and helicity dependence.
Integration with respect to $K$ can be used to eliminate the energy delta-function in \eqref{dsigma-tw}.
For fixed $E_i$, $\varkappa_i$, and $M$, the energy-momentum conservation 
fixes $K_z = k_{1z}+k_{2z}$ and, therefore, $K = \sqrt{E_K^2 - M^2 - K_z^2}$. 
Thus, the polar angle of the produced resonance is defined by 
\begin{equation}
\cos\theta_K = \fr{K_z}{\sqrt{(E_1+E_2)^2-M^2}}\,.\label{theta-K}
\end{equation}
The experiment can be repeated at different total energy $E_K = E_1+E_2$,
but, as long as $E_K$ satisfies 
\begin{equation}
E_- \le E_K \le E_+\,, \quad E_\pm = \sqrt{(\varkappa_1 \pm \varkappa_2)^2 + K_z^2 + M^2}\,,\label{E-range}
\end{equation}
the resonance with mass $M$ can be produced with a non-zero cross section.
The polar angle $\theta_K$ can also be adjusted by varying the ratio between $E_1$ and $E_2$.

Notice how different this situation is with respect to the PW collisions.
In the PW case, the (narrow) resonance production cross section is $\propto \delta(E_1+E_2-E_K)$,
and the production occurs only at the resonance.
In twisted particle annihilation, there is a finite range of energies to produce a resonance with mass $M$.
The cross section $\sigma(E_K)$ varies with $E_K$ in a periodic fashion, revealing the interference fringes
induced by the varying coefficients in front of ${\cal M}_a$ and ${\cal M}_b$ in \eqref{J2}.
Moreover, a twisted annihilation experiment running at fixed energy can {\em simultaneously} 
produce two or more resonances with close but different masses,
provided they satisfy \eqref{E-range}.
According to \eqref{theta-K}, these resonances will be emitted at different polar angles.
Thus, twisted annihilation has a built-in mass spectrometric feature.
These kinematic peculiarities will be explored in detail in the follow-up paper
\cite{kinematic}.

{\em Twisted annihilation as parity analyzer.}
To illustrate physics opportunities offered by twisted annihilation,
consider production of a hypothetical spin-0 resonance in collision of 
two twisted photons \cite{Ivanov:2019lgh}.
It can be either a pure scalar $S$, a pure pseudoscalar $P$, or their mixture.
For a pure scalar, whose effective interaction Lagrangian with photons
is described by ${\cal L}_S = g F^{\mu\nu} F_{\mu\nu} S/4$, 
one finds that the PW helicity amplitude is non-zero only for $\lambda_1=\lambda_2=\lambda$: 
${\cal M}_S = -2 g E_1E_2 \delta_{\lambda_1, \lambda_2} \bbe_{1\lambda} \bbe_{2\lambda}$. 
For Bessel photons, one obtains the twisted amplitude 
${\cal J}_S \propto {\cal J}_1 + \lambda {\cal J}_2$ with
\begin{eqnarray}
{\cal J}_1 &=& \cos(m_1\delta_1 + m_2\delta_2) \left[\cos(\delta_1 +\delta_2)(1-c_1c_2) - s_1 s_2\right]\,,
\nonumber\\
{\cal J}_2 &=& \sin(m_1\delta_1 + m_2\delta_2) \sin(\delta_1 +\delta_2) (c_1 - c_2)\,,\label{JS3}
\end{eqnarray}
where we used the short notation $c_i \equiv \cos\theta_i$, $s_i \equiv \sin\theta_i$.
The cross section explicitly depends on the photon helicities:
$\sigma_\lambda = \sigma_0 + \lambda \sigma_a$, 
where $\sigma_0 \propto {\cal J}_1^2 + {\cal J}_2^2$ is the unpolarized cross section 
and $\sigma_a \propto 2{\cal J}_1{\cal J}_2$ is the spin asymmetry.

This dependence on $\lambda$ may look surprising since the fundamental
interaction is parity-invariant.
However, unlike in the PW case, here 
we explicitly break the left-right symmetry of the initial state.
If the AM values $m_i$ are fixed, then the helicity choices $\lambda_i = \pm 1$
are not equivalent because the corresponding OAM contributions differ. 
The process would be invariant under the {\em simultaneous} sign flips 
$m_i \to -m_i$ and $\lambda_i \to - \lambda_i$,
but not with respect to $\lambda_i \to - \lambda_i$ alone.

For a pure pseudoscalar with the interaction Lagrangian 
${\cal L}_P = i g \epsilon_{\mu\nu\rho\sigma} F^{\mu\nu} F^{\rho\sigma} P/8$,
one gets ${\cal M}_P = \lambda {\cal M}_S$.
If the spin-0 particle does not possess definite parity,
one can write its PW production amplitude as ${\cal M} = a {\cal M}_S + b {\cal M}_P = (a+\lambda b){\cal M}_S$,
where the (complex) coefficients $a$ and $b$ describe the scalar/pseudoscalar contributions 
to the amplitude.
In the usual PW collision with circularly polarized photons, the cross section is
$\sigma_\lambda \propto |a|^2 + |b|^2 + 2\lambda \Re (a^* b)$.
Averaging it over the initial photon polarizations yields $\sigma_0 \propto |a|^2 + |b|^2$; 
it reveals the overall production intensity but cannot measure the amount of scalar-pseudoscalar mixing.
Twisted photons offer access to this mixing even in the unpolarized cross section. 
Using the fixed-$m$ convention for unpolarized twisted photon beams,
we obtain the twisted amplitude 
\begin{equation}
{\cal J} = (a {\cal J}_1 + b {\cal J}_2) + \lambda (b {\cal J}_1 + a {\cal J}_2)\,,
\end{equation}
with ${\cal J}_1$, ${\cal J}_2$ given in \eqref{JS3}.
Averaging $|{\cal J}|^2$ over the initial photon helicities, we get 
the unpolarized cross section
\be
\sigma_0 \propto ({\cal J}_1^2 + {\cal J}_2^2)(|a|^2 + |b|^2)
+ 4 {\cal J}_1{\cal J}_2\Re (a^* b)\,.\label{dsigma-S-P-tw}
\ee
Now both the total intensity $|a|^2 + |b|^2$ and the scalar-pseudoscalar mixing $\Re (a^* b)$
can be extracted experimentally from the energy dependence of $\sigma_0(E_K)$ 
and, specifically, from the exact location and heights of the interference fringes.
Since ${\cal J}_1$ and ${\cal J}_2$ have different dependence on $\delta_1$ and $\delta_2$,
their combinations ${\cal J}_1{\cal J}_2$ and ${\cal J}_1^2 + {\cal J}_2^2$ 
produce different interference patterns and can be distinguished.
A detailed numerical investigation of this effect with realistic twisted beams 
will be given elsewhere \cite{spin}.


{\em Twisted annihilation as spin polarizer.}
Our second example is production of a spin-1 resonance $V$ of mass $M$ 
in twisted $e^+e^-$ annihilation.
We take the PW helicity amplitude in the form 
\begin{eqnarray}\label{eq:matr.el.ep.em.anih}
{\cal M}_{\zeta_1\zeta_2\la_V}=g \bar v_{\zeta_2}(k_2)\ga_\mu u_{\zeta_1}(k_1)V^{\mu *}_{\la_V}(K)\,,
\end{eqnarray}
where $\zeta_1$, $\zeta_2$, and $\lambda_V$ are the helicities of the electron, positron, 
and the produced vector resonance, respectively, and $V^{\mu}_{\la_V}$ is the polarization vector of the spin-1 resonance.
To illustrate the main effect, we evaluate this PW helicity amplitude in the paraxial limit
$\theta_1 \to 0$, $\theta_2 \to \pi$ but with generic $\theta_K$, 
and observe that it is non-zero only for $\zeta_1 = -\zeta_2 \equiv \zeta = \pm 1/2$:
\begin{equation}
{\cal M}_{\zeta, -\zeta, \la_V} \propto e^{-i\zeta(\varphi_2+\varphi_1 - 2\varphi_K)}
\cdot (\lambda_V\cos\theta_K + 2\zeta) \,.\label{Me+e-}
\end{equation}
For twisted $e^+e^-$ annihilation, we get 
\begin{equation}
{\cal J}_{\zeta, -\zeta, \la_V} \propto (\lambda_V\cos\theta_K + 2\zeta)\, 
\cos[m_1\delta_1 + m_2\delta_2 - \zeta(\delta_1 - \delta_2)]\,,\label{Je+e-}
\end{equation}
where $m_1, m_2$ are half-integer.
Using the fixed-$m$ definition of the unpolarized electron and positron beams,
we obtain the unpolarized twisted $e^+ e^-$ cross section:
\begin{eqnarray}\label{eq:cross.sec.paraxial}
\sigma_{\la_V = \pm 1}&\propto& 
(1+\cos^2\theta_K) \nonumber \\
&&\times [1+\cos(2m_1\de_1+2m_2\de_2) \cos(\de_1-\de_2)] \\
&& +\  2\lambda_V \cos\theta_K \sin(2m_1\de_1+2m_2\de_2)\sin(\de_1 \! - \! \de_2)\,. \nonumber
\end{eqnarray}

This result shows another remarkable phenomenon, which was impossible in the PW case.
Even with unpolarized electrons and positrons,
one produces vector meson with $\lambda_V = +1$ and $\lambda_V = -1$ with unequal intensities.
The imbalance is enhanced at $\theta_K$ far from $\pi/2$
and is sensitive to the exact position at the interference fringes.
Thus, the produced vector meson {\em is} polarized on average
and its polarization can be controlled.

To give a numerical example, we consider production of the $J/\psi$ meson 
with $M=3.1$~GeV and a finite width of $\Gamma=93$~keV
in unpolarized twisted $e^+e^-$ annihilation with the following parameters:
\begin{eqnarray}\label{eq:smearingParameters}
&&E_1=1.8~\text{GeV},\ E_2=1.338~\text{GeV},\
(m_1, m_2)=(5/2, 1/2),\nonumber\\  
&&\bar\varkappa_1=0.2~\text{GeV},\ 
\bar\varkappa_2=0.1~\text{GeV},\ 
\si_i=\bar\varkappa_i/5\,.
\end{eqnarray}
Although the energies of both incoming particles are fixed, smearing over $\varkappa_i$
produces a distribution in $\theta_K$.
\begin{figure}[t]
	\centerline{
	\includegraphics[width=0.4\textwidth]{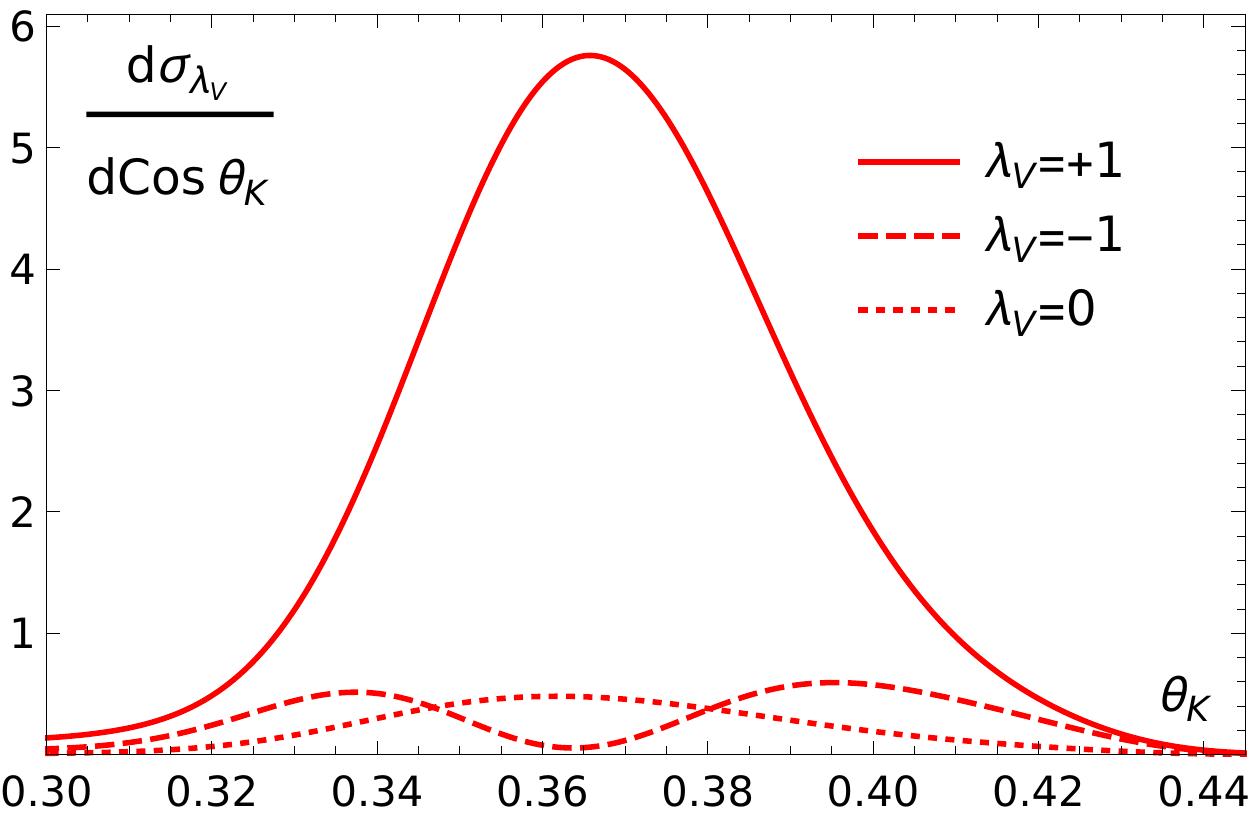}
}
	\caption{
		\label{fig:crossX4ePeM-annihilation}
Angular distribution of the $J/\psi$ meson production cross section (in arbitrary units) 
in unpolarized twisted $e^+ e^-$ annihilation 
with kinematic choice (\ref{eq:smearingParameters}).
The solid, dashed, and short-dashed lines show the cross sections for $\lambda_V = +1$, $-1$, and $0$, respectively.
	}
\end{figure}
In Fig.~\ref{fig:crossX4ePeM-annihilation} we show the resulting differential cross section $d\si_{\lambda_V}/d\cos\theta_K$ 
computed beyond the paraxial approximation for all three polarization states $\la_V=\pm 1, 0$.
With the parameter choice \eqref{eq:smearingParameters}, 
the cross section is strongly dominated by the polarization state 
$\lambda_V = +1$, with a $\approx 10\%$ admixture of the $\lambda_V = 0$ state
and even smaller contribution from $\lambda_V = -1$.

{\em Conclusions.}
In summary, we proposed a completely novel, complementary tool for doing spin physics in particle collisions.
We demonstrated that by preparing initial particles in twisted states and adjusting
their angular momenta and kinematics, 
one can access parity- and spin-dependent observables even in unpolarized inclusive cross section.
Fundamentally, it is possible because the initial twisted states explicitly break the left-right symmetry,
which leads to non-vanishing spin effects even for unpolarized cross section.
We gave two illustrations of this remarkable effect:
accessing scalar-pseudoscalar mixing of a spin-0 particle 
produced in unpolarized twisted photon collisions,
and production of polarized vector mesons in unpolarized $e^+e^-$ annihilation.
None of these effects is possible with the usual plane-wave collisions.

Experimental exploration of these phenomena requires significant development of accelerator instrumentation
to make high-energy physics with twisted particles possible.
We believe that the novel opportunities in hadronic physics offered by twisted particles
present a compelling scientific case to justify this dedicated work.

\section*{Acknowledgments}
I.P.I. thanks the Institute of Modern Physics, Lanzhou, China, for financial support and hospitality during his stay.
I.P.I. acknowledges funding from the Portuguese
\textit{Fun\-da\-\c{c}\~{a}o para a Ci\^{e}ncia e a Tecnologia} (FCT) through the FCT Investigator 
contract IF/00989/2014/CP1214/CT0004 and project PTDC/FIS-PAR/29436/2017,
which are partially funded through POCI, COMPETE, Quadro de Refer\^{e}ncia
Estrat\'{e}gica Nacional (QREN), and the European Union.
I.P.I. also acknowledges the support from National Science Center, Poland,
via the project Harmonia (UMO-2015/18/M/ST2/00518).
P.M.Z. is supported by the National Natural Science Foundation of China (Grant No. 11975320).
A.V.P. and N.K. thank the Chinese Academy of Sciences President's International Fellowship Initiative for the support
via grants No. 2019PM0036 (A.V.P.) and No. 2017PM0043 (N.K.).

\bibliography{res-short-v1.6}
\bibliographystyle{apsrev4-1}

\newpage
\appendix
\section*{Supplementary information}

\subsection{Description of twisted photons and electrons}

We describe twisted photons with the formalism developed in 
\cite{Jentschura:2010ap,Jentschura:2011ih,Knyazev:2019}.
For definiteness, we work in the Coulomb gauge.
A monochromatic plane-wave electromagnetic field with helicity 
$\lambda = \pm 1$ is described by
\be
\bA_{\bk \lambda}(\br) = \bbe_{\bk \lambda}\, e^{i\bk \br}\,.\label{PW1}
\ee
The polarization vector is orthogonal to the wave vector: $\bbe_{\bk\lambda} \bk = 0$.
To construct a Bessel twisted photon, we fix a reference frame, select an axis $z$, and 
write it as a superposition of PW photons with fixed longitudinal momentum 
$k_z = |\bk|\cos\theta$, fixed modulus of the transverse momentum $\varkappa = |\bk_\perp| = k\sin\theta$,
but arriving from different azimuthal angles $\varphi_k$.
A twisted photon with a definite $z$-projection of the {\em total} AM $m$
and definite helicity $\lambda = \pm 1$ is written in Eqs.~\eqref{tw1-ph} and \eqref{a-ph}.
The usual dispersion relation holds for every plane wave component: $k_z^2 + \varkappa^2 = E^2$.

The Fourier amplitude \eqref{a-ph} is an eigenstate not only of the $z$-component of the total AM operator $\hat{J}_z$
but also the OAM operator $\hat{L}_z = -i \partial/\partial \varphi_k$.
For a scalar twisted particle, it would imply that such a state possesses a well-defined OAM.
However, this property is not shared by $\bbe_{\bk \lambda}$,
which is an eigenstate of $\hat{J}_z$ with a zero eigenvalue but not of $\hat{L}_z$ or $\hat{s}_z$ separately.
This is a manifestation of the spin-orbital interaction of light,
which gives rise to a variety of remarkable optical phenomena \cite{Bliokh:2015yhi}.
Nevertheless, in the paraxial approximation $\varkappa/|k_z| = |\tan\theta| \ll 1$, 
the spin-orbital coupling is suppressed and one can deal with approximately conserved 
$s_z \approx \lambda$ and $\ell = m - \lambda$.

Each PW component of a twisted photon
contains its polarization vector $\bbe_{\bk \lambda}$, which is orthogonal 
to the momentum of that particular PW component: $\bbe_{\bk\lambda} \bk = 0$.
To describe it in the chosen coordinate frame, we first define the eigenvectors ${\bm \chi}_\sigma$, $\sigma= \pm 1, 0$,
of the helicity operator $\hat{s}_z$ defined with respect to the fixed axis $z$: 
$\hat{s}_z {\bm \chi}_\sigma = \sigma {\bm \chi}_\sigma$.
Their explicit form is 
\be
{\bm \chi}_{0}=
\left(
\begin{tabular}{c}
	0 \\
	0 \\
	1 \\
\end{tabular}
\right),\ 
{\bm \chi}_{\pm 1}= \fr{\mp 1}{\sqrt{2}}
\left(
\begin{tabular}{c}
	1 \\
	$\pm i$ \\
	0 \\
\end{tabular}
\right)\,, \quad {\bm \chi}^*_\sigma {\bm \chi}_{\sigma^\prime} = \delta_{\sigma\sigma^\prime}\,.
\label{chi}
\ee
The polarization vector can be expanded in the basis of ${\bm \chi}_\sigma$:
\be
\bbe_{\bk \lambda}=
\sum_{\sigma=0,\pm 1} e^{-i\sigma \varphi_k}\,
d^{1}_{\sigma \lambda}(\theta)  \,\bm \chi_{\sigma}\,.
\label{bbe-chi}
\ee
The explicit expressions for Wigner's $d$-functions~\cite{LL3} are:
\be
d^{1}_{\sigma\lambda} = \mmmatrix{\cos^2\fr{\theta}{2}}{-\fr{1}{\sqrt{2}}\sin\theta}{\sin^2\fr{\theta}{2}}%
{\fr{1}{\sqrt{2}}\sin\theta}{\cos\theta}{-\fr{1}{\sqrt{2}}\sin\theta}%
{\sin^2\fr{\theta}{2}}{\fr{1}{\sqrt{2}}\sin\theta}{\cos^2\fr{\theta}{2}}%
\ee
The first, second, and third rows and columns of this matrix correspond to the indices $+1,\, 0,\, -1$.
Performing the summation in Eq.~\eqref{bbe-chi}, one gets explicit expressions for the polarization vectors:
\be
\bbe_{\bk \lambda}= \fr{\lambda }{\sqrt{2}}\triplet{-\cos\theta\cos\varphi_k + i \lambda \sin\varphi_k}%
{-\cos\theta\sin\varphi_k - i \lambda \cos\varphi_k}{\sin\theta}\,,\quad \lambda= \pm 1\,.\label{bbe-explicit}
\ee
Finally, when describing a counter-propagating twisted photon defined in the same reference frame
with respect to the same axis $z$, one can use the above expressions assuming that $k_z < 0$
and replacing $m \to - m$ in the Fourier amplitude \eqref{a-ph}.
The expression for the polarization vector \eqref{bbe-explicit} stays unchanged,
but $\cos\theta < 0$. 

Twisted states have been experimentally demonstrated not only for photons
but also for electrons \cite{Uchida:2010,Verbeeck:2010,McMorran:2011}. 
To describe them in a fully relativistic manner, 
we use the definitions of \cite{Serbo:2015kia,Bliokh:2017uvr};
other works, such as \cite{Bliokh:2011fi,Karlovets:2012eu}, use slightly different conventions.
The PW electron with energy $E$, momentum $\bk$, and helicity $\zeta = \pm 1/2$ is described by
$e^{i\bk\br}u_{\zeta}(k)/\sqrt{2E}$ with the bispinor
\be
u_{\zeta}(k) = \doublet{\sqrt{E+m_e}\,w^{(\zeta)}}{2 \zeta \sqrt{E-m_e}\,w^{(\zeta)}}\,, 
\ee
where 
\be
w^{(+\frac{1}{2})} = \doublet{\cos\frac{\theta}{2}\, e^{-\frac{i\varphi_k}{2}}}{\sin\frac{\theta}{2}\, e^{\frac{i\varphi_k}{2}}},
\ w^{(-\frac{1}{2})} = \doublet{-\sin\frac{\theta}{2}\, e^{-\frac{i\varphi_k}{2}}}{\cos\frac{\theta}{2}\,e^{\frac{i\varphi_k}{2}}}\,.\label{PWspinors}
\ee
The bispinors are normalized as 
$\bar u_{\zeta_1}(k) u_{\zeta_2}(k) = 2m_e\, \delta_{\zeta_1, \zeta_2}$.
The negative-frequency solutions of the Dirac equation are constructed as
\be
v_{\zeta}(k) = \doublet{-\sqrt{E-m_e}\,w^{(-\zeta)}}{2\zeta \sqrt{E+m_e}\,w^{(-\zeta)}}\,,
\ee
with the same spinors $w$ as in \eqref{PWspinors}.
We use this basis of plane-wave solutions of the Dirac equation to construct 
the Bessel twisted state of the electron in Eq.~(\ref{bessel}).
The similar expression holds for the negative-frequency solutions.
For the sake of simplicity, we performed calculations of the twisted $e^+e^-$ annihilation
in the massless limit $m_e \to 0$. Restoring the finite mass of the electron 
does not change the results in any noticeable way.

\subsection{Calculating twisted particle annihilation}

In the PW case, the $S$-matrix amplitude of the $2\to 1$ process has the form
\begin{equation}
S_{PW} = i(2\pi)^4 \delta^{(4)}(k_1 + k_2 - K)  {{\cal M}(k_1,k_2;K) \over \sqrt{8 E_1 E_2 E_K}}\,.
\label{SPW}
\end{equation}
Here, ${\cal M}(k_1,k_2;K)$ is the plane-wave invariant amplitude 
calculated according to the standard Feynman rules.
Squaring this amplitude, regularizing the squares of delta-functions, and diving by flux, 
as described, for instance, in \cite{LL4}, 
one gets the cross section 
\bea
d\sigma &=& {\pi \delta(\Sigma E) \over 4 E_1 E_2 E_K v} |{\cal M}|^2  \, 
\delta^{(3)}(\bk_1 + \bk_2 - \bK) \, d^3 K\,,\nonumber\\[2mm]
\sigma &=& {\pi \delta(\Sigma E) \over 4 E_1 E_2 E_K v} |{\cal M}|^2\,,\label{sigma-PW-0}
\eea
where $\delta(\Sigma E) \equiv \delta(E_1+E_2-E_K)$.
As should be expected in the PW case, the final momentum is fixed at $\bK = \bk_1 + \bk_2$ 
and the dependence on the total energy of the colliding particles is proportional to $\delta(E_1+E_2-E_K)$.
The production process occurs only when the initial particles are directly ``at the resonance''.

Let us now consider collision of two Bessel states or arbitrary particles 
$|\varkappa_1,m_1\rangle$ and $|\varkappa_2,m_2\rangle$
which are defined in the same reference frame and with respect to the same axis $z$.
The final particle with mass $M$ is still described in the basis of plane waves,
and its momentum $\bK$ and energy $E_K$ satisfy $E_K^2 = M^2 + \bK^2$.
We follow the procedure of \cite{Jentschura:2010ap,Ivanov:2011kk,Ivanov:2016jzt},
which adapts the general theory of scattering of non-monochromatic, arbitrarily shaped,
partially coherent beams developed in \cite{Kotkin:1992bj} to the collisions of Bessel twisted states.
The $S$-matrix element of this process is 
\be
S = \int {d^2 \bk_{1\perp} \over (2\pi)^2} {d^2 \bk_{2\perp} \over (2\pi)^2} 
a_{\varkappa_1 m_1}(\bk_{1\perp}) a_{\varkappa_2, -m_2}(\bk_{2\perp}) S_{PW}\,.\label{S-tw}
\ee
The negative sign in front of $m_2$ reflects the fact that the second particle propagates on average
in the $-z$ direction.
Substituting in \eqref{S-tw} the Fourier amplitudes of the Bessel states,
we get
\be
S = i (2\pi)^4 \fr{\delta(\Sigma E) \delta(\Sigma k_z)}{\sqrt{8 E_1 E_2 E_K}} 
{(-i)^{m_1-m_2} \over (2\pi)^3\sqrt{\varkappa_1\varkappa_2}} \cdot {\cal J}\,,\label{S-tw2}
\ee
where $\delta(\Sigma k_z) \equiv \delta(k_{1z}+k_{2z}-K_z)$.
The twisted amplitude ${\cal J}$ is defined in \eqref{J}.
Since it contains the equal number of integrations and delta-functions, it can be calculated exactly \cite{Ivanov:2011kk}. 
It is non-zero only if the moduli of the transverse momenta $\varkappa_i = |\bk_{i\perp}|$ 
and $K \equiv |\bK_\perp|$ satisfy the triangle inequalities \eqref{ring} and form a triangle with the area
\be
\Delta 
= {1 \over 4} \sqrt{2 K^2\varkappa_1^2 + 2 K^2\varkappa_2^2 + 2\varkappa_1^2\varkappa_2^2 
	- K^4 - \varkappa_1^4 - \varkappa_2^4}\,.\label{area}
\ee
Out of many plane wave components ``stored'' in the initial twisted particles, 
the integral \eqref{J} receives contributions from exactly two plane wave combinations shown in Fig.~\ref{fig-2configurations}
with the following azimuthal angles:
\begin{eqnarray}
\text{configuration a:} &\quad& \varphi_1 = \varphi_{K} + \delta_1\,,\quad \varphi_2 = \varphi_{K} - \delta_2\,,\nonumber\\
\text{configuration b:} &\quad& \varphi_1 = \varphi_{K} - \delta_1\,,\quad \varphi_2 = \varphi_{K} + \delta_2\,. \nonumber
\end{eqnarray} 
\hskip 0mm

\noindent The inner angles of the triangle $\delta_1$, $\delta_2$ are determined by 
\be
\cos\delta_1 = {\varkappa_1^2 + K^{2} - \varkappa_2^2 \over 2\varkappa_1 K},
\quad 
\cos\delta_2 = {\varkappa_2^2 + K^{2} - \varkappa_1^2 \over 2\varkappa_2 K}.
\label{delta_i}
\ee
As a result, ${\cal J}$ can be calculated exactly \cite{Ivanov:2011kk}: 
\begin{eqnarray}
{\cal J} = e^{i(m_1 - m_2)\varphi_{K}}{\varkappa_1 \varkappa_2 \over 2\Delta}
\Bigl[{\cal M}_{a}\, e^{i (m_1 \delta_1 + m_2 \delta_2)} \nonumber\\
\quad \quad \quad +\ {\cal M}_{b}\, e^{-i (m_1 \delta_1 + m_2 \delta_2)}\Bigr]\,.\label{J3}
\end{eqnarray}
This expression displays the hallmark feature of the twisted particle collision processes:
the interference between two PW amplitudes ${\cal M}_{a}$ and ${\cal M}_{b}$ 
calculated for the two distinct initial PW pairs shown in Fig.~\ref{fig-2configurations} 
but the same final momentum $\bK_\perp$.
They represent two distinct paths, in momentum space, to arrive at the same final state from the initial twisted states \cite{Ivanov:2016jzt}.

When presenting numerical results, we use not the pure Bessel states, which are not normalizable in the transverse plane,
but the monochromatic $\varkappa$-smeared wave packets given in Eq.~\eqref{WP}.
Instead of repeating the derivation of the cross section, one just applies this smearing procedure
with the functions $f_1(\varkappa_1)$ and $f_2(\varkappa_2)$ to $S$-matrix amplitude \eqref{S-tw2}.
Therefore, ${\cal J} \cdot \delta(k_{1z}+k_{2z}-K_z)$ for the pure Bessel states turns now into 
\begin{equation}
\lr{{\cal J}} = \int\limits_0^{E_1} \! d\varkappa_1 \! \int\limits_0^{E_2} d\varkappa_2 f_1(\varkappa_1) f_2(\varkappa_2) \delta(k_{1z} + k_{2z} - K_z) 
{{\cal J} \over \sqrt{\varkappa_1\varkappa_2}}\,.\label{J-smeared}
\end{equation}
In numerical calculations, we use the Gaussian smearing functions of the following form:
\begin{eqnarray}
f_i(\varkappa_i)=n_{i}
\sqrt{\varkappa_i}
\exp\left[
-\frac{(\varkappa_i-\bar\varkappa_i)^2}{2\si_i^2}\right]\,.
\end{eqnarray}
The normalization condition $\int_0^{E_i} d\varkappa |f_i(\varkappa)|^2=1$
fixes the normalization constants $n_i$.
The differential cross section \eqref{dsigma-tw} turns into
\begin{equation}
d\sigma \propto |\lr{{\cal J}}|^2 \delta(E_1+E_2-E_K) \, d^3K\,.\label{dsigma-tw-2}
\end{equation}
Removing the energy delta-function, one obtains a non-trivial angular distribution over a finite range
of polar angles:
\begin{equation}
d\sigma \propto E_K^2 \beta_K\, |\lr{{\cal J}}|^2\, d\Omega_K\,.\label{dsigma-tw-3}
\end{equation}

\end{document}